\documentclass[
    ,final            
  ]
  {aipproc}
\usepackage{graphicx} 
\layoutstyle{8x11double}

\def\xte{XTE\,J1810-197}

\def\uu {4U\,0142$+$614}
\def\ee {1E\,1048.1$-$5937}

\def\ea {1E\,2259$+$586}

\def\cxou{CXOU J164710.2$-$455216}

\def\swift{{\em Swift}}

\def\XMM{{\em XMM-Newton}}
\def\CXO{{\em Chandra}}

\def\ergscm2{\rm erg\,cm^{-2}\,s^{-1}}
\def\ergs{\rm erg\,s^{-1}}

\begin{document}

\title[Transient Phenomena in Anomalous X--ray Pulsars]{Transient Phenomena in
Anomalous X--ray Pulsars}


\author{G.L. Israel}{address={INAF -
Astronomical Observatory of Roma, Via
Frascati 33, 00040 Monte Porzio Catone, Italy}}

\author{F. Bernardini}{address={University of Rome ``La Sapienza'',
Department of Physics, P.le A. Moro 5, 00185, Roma, Italy}}

\author{M. Burgay}{address={INAF - Astronomical Observatory of 
Cagliari, localit\'a Poggio dei Pini, strada 54, 09012 Cagliari, Italy}}

\author{N. Rea}{address={SRON Netherlands Institute for Space Research,
Sorbonnelaan, 2, 3584CA, Utrecht, The Netherlands}}

\author{A. Possenti}{address={INAF - Astronomical Observatory of 
Cagliari, localit\'a Poggio dei Pini, strada 54, 09012 Cagliari, Italy}}

\author{S. Dall'Osso}{address={INAF -
Astronomical Observatory of Roma, Via
Frascati 33, 00040 Monte Porzio Catone, Italy}}

\author{L. Stella}{{address={INAF -
Astronomical Observatory of Roma, Via
Frascati 33, 00040 Monte Porzio Catone, Italy}},altaddress={on behalf
of a larger team (the complete list is reported in the 
acknowledgment section)}}

%

\begin{abstract}
In 2003 a previously unpulsed Einstein and ROSAT source cataloged
as soft and dim (L$_X$ of few $\times$10$^{33}\ergs$) thermal emitting object,
namely \xte, was identified as the first unambiguous transient Anomalous X-ray
Pulsar. Two years later this source was also found to be a bright highly
polarized transient radio pulsar, a unique property among both AXPs and radio
pulsars. In September 2006 \swift\ Burst Alert Telescope (BAT) detected an
intense burst from the candidate AXP \cxou, which entered in an outburst state
reaching a peak emission of at least a factor of 300 higher than quiescence. 
Here, we briefly outline the recent results concerning the outburst phenomena
observed in these two AXPs. In particular, \xte\ has probed to be a unique
laboratory to monitor the timing and spectral properties of a cooling/fading
AXP, while new important information have been inferred from X-ray and radio
band simultaneous observations. \cxou\ has been
monitored in X-rays and radio bands since the very beginning of its 
outbursting state allowing us to cover the first phases of the
outburst and to study the timing and spectral behavior during the first
months. 
\end{abstract}


\maketitle


\section{Introduction}
At the beginning of the X--ray astronomy era the study of the X-ray variability
(both of the flux and timing properties) of Cen X--3 allowed
astronomers to
unambiguously assess the binary nature of the source and to identify the
accretion of mass, flowing from the companion onto a rotating neutron star, as
the main mechanism to produce X--rays \citep{cenx-3a,cenx-3b}. Since then,
several classes of
high energy sources hosting a compact object have been identified and
their variability, observed on different timescales, used to test
models and/or to study the emission mechanisms as a function of flux (while
keeping fixed other parameters, such
as the distance and the geometry of the system). For the class of accreting
neutron stars this approach was successfull in confirming, among others, the
presence of a centrifugal barrier to accretion, testing the dependence of the
period derivative with respect to the source luminosity, and studying the change
in the photon propagation direction as a function of the accretion rate by
means of the pulse shape changes 
\citep[fan  and pencil beam model; as an example see ][]{exo1,exo2,lorella89}. 
Isolated neutron stars are relatively constant sources, making
the above approach unreliable. However, there are two small classes of isolated
neutron stars which show spectacular events
during which their luminosity may change up to 10 orders of magnitude
on timescales down to few milliseconds. These objects are better known as
Anomalous X--ray Pulsars (AXPs; 10 objects plus 1 candidate) and 
Soft $\gamma$-ray Repeaters \citep[SGRs; 4 objects plus 2 candidates; for a
review see][]{wot06}. It is believed that AXPs and SGRs are linked at some
level, owing to their similar timing properties (spin periods in the 2-12\,s
range and  period derivatives \.P in the  10$^{-13}$--10$^{-11}$ s\,s$^{-1}$
range). Both classes  have been proposed to host neutron stars whose emission is
powered by the decay of their extremely strong magnetic fields 
\citep[$>10^{15}$\,G;][]{DT92, TD95}.
 
Different types of  X-ray flux variability have been displayed by AXPs. From
slow and moderate flux changes (up to a factor of few) on timescales of years
(virtually all the object of the class),  to moderate-intense outbursts
(flux variations of a factor up to 10) lasting for 1-3 years (\ea,
and \ee),  to dramatic and intense SGR-like burst activity (fluence of
$10^{36}-10^{37}$ ergs) on sub-second timescales \citep[\uu, \xte, \ea\
and \ee; see ][for a review on the X-ray variability]{2007London}. 
The first notable recorded case of flux variability was the
2002 bursting/outbursting event detected from \ea, the only known event in which
a factor of $\sim$10 persistent flux enhancement in an AXP was followed
(or proceeded) by the
onset of a bursting activity phase during which the source displayed more than
80 short bursts \cite{fotis04,pete04}. The timing and spectral properties of the
sources changed significantly and recovered the pre-bursting activity phase
values within few days, likely due to the relatively high luminosity DC level
($\sim 10^{35}\ergs$). However, it was only in 2003 that the first transient AXP
was discovered, namely \xte, which displayed a factor of $>$100  persistent
flux enhancement with respect to the unpulsed pre-outburst quiescent luminosity
level \citep[$\sim 10^{33}\ergs$; ][]{alaa04,eric04,ioxtej1810,nanda04}.
Unfortunately, the initial phases of the outburst were missed and we do not know
whether a bursting activity phase, similar to that of \ea, occurred also for
this source, though four bursts have been detected by RossiXTE between 2003
September and 2004 April and unambiguously associated with \xte\ \cite{pete05}.
In 2006 the candidate AXP \cxou\ displayed a bursting-outbursting
behavior with a maximum flux variability of $>300$, followed by extreme
and daily changes both in the spectral and timing properties
\cite{mike06a,mike06b,iocxou}.
\begin{figure}[thb]
  \includegraphics[width=1.\textwidth]{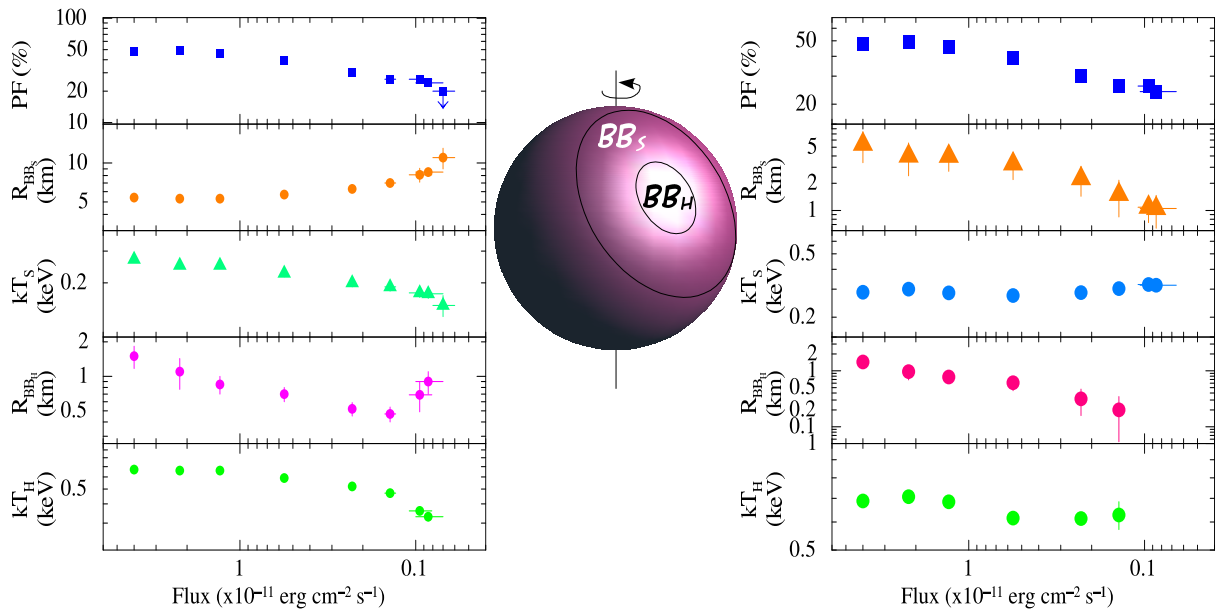}
  \caption{Evolution of the spectral parameters for the
2BB (left panel) and 3BB (right panel) models together with the pulsed fraction
as a function of the observed 1-10\,keV band flux. Central Panel: schematic
view of the assumed scenario with marked the emission regions of the hottest BB
components.}
\label{fig:XTE}
\end{figure}

These two sources currently represent our best opportunity in order to study the
evolution of the main spectral and timing parameters as a function of flux by
keeping fixed other parameters (such as distance and geometrical angles)
otherwise difficult to infer (similarly to the pioneering studies on accreting
X--ray pulsars). In the following pages we will briefly outline (i) the
recent results obtained from the analysis of the latest 4 years \XMM\ monitoring
observations of \xte\ as it approached to quiescence, (ii) the
comparison of X-ray and radio emission from \xte\ by means of two
$\sim$8hr-long simultaneous campaigns with \XMM\ and Parkes, (iii) and the
results of the first 6 months monitoring of \cxou\ during the first phases of 
its current outburst by means of \swift, \CXO\ and \XMM\ in the X-rays and
Parkes in the radio band.

\section{XTE J1810--197: from outburst to quiescence}

Since the very first \XMM\ 2003 observations of \xte, carried out
approximately one year after the onset of the outburst, it was evident
\cite{eric04} that the source spectral shape (two blackbodies with
kT=0.29$\pm$0.03\,keV and R$_{BB}\approx$5.5\,km, and
kT=0.70$\pm$0.02\,keV and R$_{BB}\approx$1.5\,km;
L$_X\sim$5$\times$10$^{34}\ergs$ in the 0.5-10\,keV range) was significantly
different from that serendipitously recorded by ROSAT in 1992 (one BB with
kT$\approx$160\,eV and R$_{BB}\approx$10\,km; extrapolated luminosity in the
0.5-10\,keV range of L$_X\sim$7$\times$10$^{32}\ergs$ and  for a distance of
3.3\,kpc). Moreover, the source
showed a 5.54\,s pulsation with a pulsed fraction of nearly 45\% during
outburst, while an upper limit of 24\% was inferred from the ROSAT data. The
above issues originated a number of important questions awaiting for an answer: 
Is the soft BB component detected by \XMM\ evolving into the quiescent
BB component seen by ROSAT ? Alternatively, is the emission from the 
whole surface always present ? What happens to the higher temperature BB
component as the source approaches to quiescence ? Which is
the pulsed fraction level of the source in quiescence (if detectable) ? Does the
outburst changed permanently the timing/spectral properties (such as the pulsed
fraction, the flux and temperature or size of the quiescent BB component) of the
source ? 

In order to try answering to the above questions we reduced all the
archival (6) and still proprietary (2) \XMM\ observations \citep[for the
details see][]{nanda07,fede07} and fitted the eight spectra all together.
All the spectra have been rebinned in order to ensure that each
background-subtracted spectral channel has at least 25 counts, and that the
instrumental energy resolution is not oversampled by a factor larger than 3
\citep[][; indeed the correct application of the above rules prevents 
artificially low (good) reduced $\chi^2$s.]{xmmrebin07}. 
In particular, we can outline the obtained results as follows:

{\bf 2BB model: } By extending the spectral recipe outlined by
\cite{eric04,eric05} we applied the two BB spectral fit analysis to the fading
phases of \xte\ until March 2007 when the flux source was $\sim$1.2 times above
the pre-outburst level (reduced $\chi^2\sim$1.23 for 975 degree of freedom,
d.o.f.; N$_H$=0.58$\pm$0.02\,$\times$10$^{22}$\,cm$^{-2}$). While the soft BB
component smoothly approaches
to that in quiescence (see Figure \ref{fig:XTE}, left panel, 2$^{nd}$ and
3$^{rd}$ plots), we note a number of ambiguities difficult to account for by
means of simple assumptions. The hard component BB radius is not monotone and it
increases after 2.5 years of smooth decrease (left panel, 4$^{th}$ plot) while
the temperature approaches to that of the soft BB in 2003 (left panel, 5$^{th}$
plot). Moreover, none of the spectral parameters or components is able to
account for the flattening, at the 25\% level, showed by the pulsed fraction
evolution (left panel, 1$^{st}$ plot) \cite{fede07}. 

{\bf 3BB model: } The addition of a further BB component gives a better  fit 
(reduced $\chi^2\sim$1.15 for 973 d.o.f.;
N$_H$=0.70$\pm$0.02\,$\times$10$^{22}$\,cm$^{-2}$;
F-test probability gives 7.3$\sigma$) though not yet satisfactory. Notably, the
fit gives parameters and
flux (for the coldest BB) which are virtually equal to those inferred in
quiescence. Even more interesting, the hottest BB components show a nearly
constant evolution of the temperatures (see Figure\,\ref{fig:XTE}, right panel,
3$^{rd}$ ad 5$^{th}$ plots), leaving the radii as the only variable parameters
to account for the decaying phases of the outburst (right panel,
2$^{nd}$ and 4$^{th}$ plots). Since September 2006 the hottest BB is not 
anymore needed to fit the spectra (upper limit of
$\sim$5$\times$10$^{-15}\ergscm2$). In this scenario, the already mentioned
flattening of the pulsed fraction might be easily accounted for by the
disappearance 
of the hot BB. A pulse phase spectroscopic analysis do not show any phase 
lag between the two highest temperature BBs \cite{fede07}. 

{\bf Further components: } During the first two XMM observations (2003-2004)
the spectral fit residuals clearly suggest (at a 3.2$\sigma$ confidence level)
the likely presence of an additional hard component above 7--8\,keV which we are
not able to characterize due to poor statistics in this band. We can only
speculate that might be somewhat related to the presence of a hard
power-law-like component detected in other AXPs \cite{lucien04,lucien06}
by INTEGRAL and which extends up to (at least) 200keV \cite{diego06};
\begin{figure}
  \includegraphics[width=.48\textwidth]{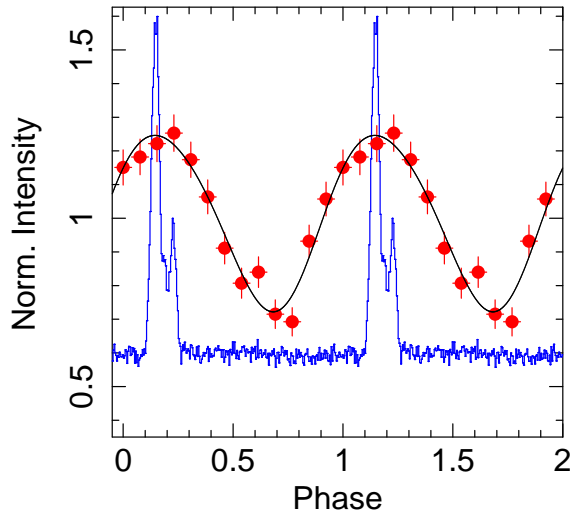}
  \caption{\xte\ \XMM\ PN+MOS 0.5-10keV and Parkes 20cm light curves, folded to
the spin period, carried out during the September 2006 campaign. Superimposed
to the X--ray folded light curve is the best sinusoidal fit \citep[adapted
from][]{io07_xradio}.}
\label{fig:XR_xtej1810}
\end{figure}

{\bf X--ray and radio campaigns: } The simultaneous radio and X--ray
observations of \xte\ carried out in September 2006 and March 2007 showed that
the pulse alignment between the two bands is high and stable (see
Figure\,\ref{fig:XR_xtej1810}), while the pulse width is relatively small
($\sim$0.1 in phase) in the radio, though not unusual among radio pulsars
\citep{io07_xradio,lorimer06}. This suggests that the X--ray and radio
emitting ragions are likely different but nearby (or superimposed), the X--rays
likely
coming from a larger area. Moreover, during the
first campaign large radio flux
($\sim$50\%) and pulse shape variations have been detected which do not
correlate with any change (at a few percent level) of the X--ray timing and/or
spectral parameters. This suggests that the X--ray emission likely originates
deep in the crust and, more in general, the radio and X--ray mechanisms appear 
different. 

\begin{figure}[bth]
  \includegraphics[width=1.\textwidth]{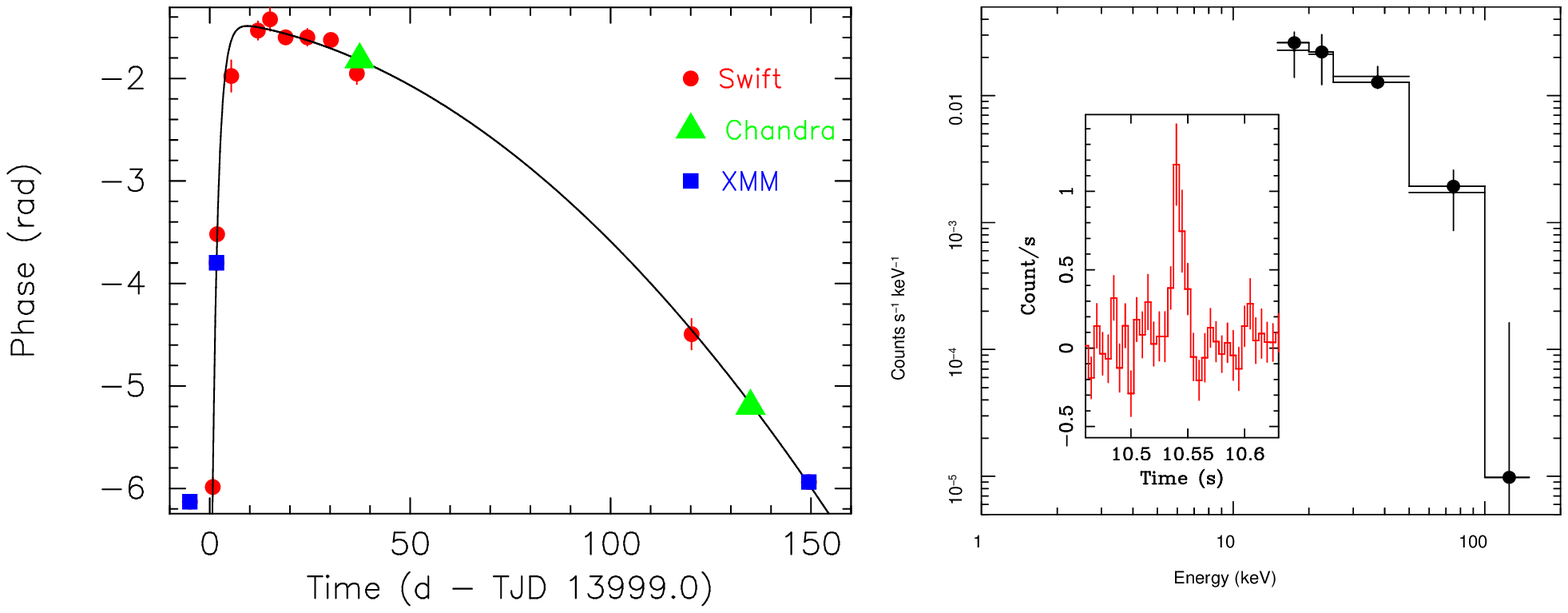}
  \caption{Left Panel: phases of the \swift\ XRT,  \XMM\ and
\CXO\ observations of \cxou: a large and quick decaying component is clearly
present.  Note that the first \XMM\ point (at day -5)  would be at the reported
phase only in the hypothesis that the pre- and post-glitch parameters are
similar \citep[for more details see][]{ioxtej1810}. Right Panel:
the 15-150\,keV \swift\ BAT spectrum of the 20ms long burst detected from \cxou\
on 21st September 2006, together with the 5ms-binned BAT lightcurve in the time
interval around the trigger (inset).}
\label{fig:cxou}
\end{figure}

We emphasize that the spectral model used above (three BBs) is a first
attempt to infer the evolution of a number of physical quantities while making
use of well assessed and reliable components. The three BB model has the
advantage of being model-independent, and of minimizing the number of variable
parameters during the outburst (only the radii are changing and both
decreasing). With respect to other model recently developed \citep[see for
example][]{tolga07} we carried out a cross check with the timing properties and
rejected all those models/components not able to reproduce the pulsed fraction
evolution. 
All the above inferred information are important in the effort we are currently
making in developing and using more detailed and complex (but necessarily
model-dependent) modelizations. In particular, the simultaneous fit of the
spectra and energy resolved folded light curves might provide a tool to 
infer the geometry of the surface temperature distribution and to independently
check the goodness of the assumed spectral model(s). Finally, we note that the
three BB model fit implies that the BB component accounting for the emission
from the whole sourface is almost unpulsed (pulsed fraction of 9\%$\pm$2\%);
this prediction can be easily checked once the source will return into the
quiescent state.

\section{CXOU J164710.2--455216: from quiescence to outburst}

On 2006 September 21, the candidate AXP \cxou\ emitted an intense
($\sim$10$^{39}\ergs$) and short (20ms) burst promptly detected by the \swift\
BAT.  Together with the burst, large changes in the timing and spectral
properties of the persistent component were detected and seen evolving during
the subsequent weeks. In particular, the \swift\ XRT monitoring (plus two
proprietary \XMM\  and two  archival \CXO\ observations ) during the first six
months since the outburst allowed to infer the following characteristics: 


{\bf The BAT burst: } The prompt event recorded by \swift\ BAT has an
exponential time decay $\tau$ of 3.3$\pm$1.0\,ms (1$\sigma$ confidence level)
and the spectrum can be fitted with both a blackbody with $kT$ of
9.9$\pm^{2.8}_{2.2}$ keV and a $\Gamma$ of 1.8
$\pm{0.5}$ (see Figure\,\ref{fig:cxou}, right panel). In both cases a fluence of
$\approx 10^{-8}$\,erg\,cm$^{-2}$ corresponding to a total energy of  $\sim 2
\times 10^{37}$\,ergs (for a fiducial distance of 5kpc).   Compared with the
properties of the previously detected AXP bursts, the current burst has a
duration within 1$\sigma$  from the log-normal distribution average value
inferred for \ea, while the fluence is significantly (a factor of about 50)
larger than the mean \cite{fotis04}.  There is only one burst, out of
$\sim$80 detected from \ea, with duration and fluence comparable with that of
\cxou, while a total of three bursts (over the whole burst duration
distribution) have fluence comparable or slightly larger than that of BAT. 

{\bf The phase-coherent timing and the glitch: } The pulse phase evolution 
is consistent with the occurrence of a large
glitch ($\Delta \nu / \nu \sim 10^{-4}$),  the largest ever detected from a
neutron star\footnote{The glitch
detection was obtained by minimizing the number of variable peaks 
in the pulse profile. A less significant timing solution
is feasible and requesting only a \.P component. We consider the latter
unlike since it would imply high variability for all the peaks; see
\cite{iocxou} for details.}. We also detected a quadratic component in the pulse
phases 
corresponding to a $\dot{\rm P}=9.2(4) \times 10^{-13}$ s s$^{-1}$ and
implying a magnetic field strength of 10$^{14}$ G (see
Figure\,\ref{fig:cxou}, left panel). The first 1-10\,keV \swift\ XRT spectrum
was carried out $\sim$13 hours after the burst detection and showed, in addition
to a $kT \sim0.65$\,keV blackbody ($R_{BB}\sim1.5$\,km) , a $\Gamma\sim2.3$
power-law component accounting for about 50\% of the observed flux
(alternatively, two blackbodies with $kT_{s}=0.50\pm 0.05$\,keV with
$R_{BBs}=3.2\pm0.4$\,km, and
$kT_{h}=1.1\pm_{0.1}^{0.2}$\,keV with $R_{BBh}=0.5\pm0.1$\,km; all the
uncertainties are at the 90\% level);

{\bf The flux and pulsed fraction evolution: } The flux decay of \cxou\ is
well
described by the function $F \propto t^{\alpha}$, with index ${\alpha}$ of
--0.28$\pm$0.05 (similar to the case of the 2002 \ea\ burst-active phase).
Moreover, we found that the PL component decays more rapidly (index ${\alpha}$
of --0.38$\pm$0.11; 90\% uncertainty) than the BB flux (index ${\alpha}$ 
of --0.14$\pm$0.10). The pulsed fraction of the 10.61\,s pulsations was seen to
drop from a value of $\sim80\%$ (as recorded by an \XMM\ observation few days
before the burst) to $\sim10\%$ few hours after the BAT event. The spectral and
timing analysis clearly show that only the blackbody component is responsible
for the pulsed flux. In particular, the signal fractional rms  as a function of
time is well fitted by a power-law with index ${\alpha}$  of +0.38$\pm$0.11,
equal (but with opposite sign) to the power-law component decay. 

{\bf The quiescent properties: } Archival \CXO\ data analysis revealed that the
modulation in quiescence is 100\% pulsed at energies above $\sim$4keV and
consistent with the (unusually small-sized) blackbody component being occulted
by the neutron star as it rotates.

{\bf Radio observations: } Since the onset of the outburst, \cxou\ has been 
routinely observed in the radio band from Parkes. No pulsed radio emission has
been detected so far with an upper limit of 0.04$mJy$ at 1400 MHz
\cite{marta06}.


All these results confirmed unambiguously that \cxou\ is a transient and
bursting AXP, showing an unusually high pulsed fraction level in quiescence.
In particular, the comparison of the cumulative properties of \cxou\ with those
of other AXPs which showed a similar behavior in the latest years confirmed
that these outbursting events are more common than previously thought.The
way and timescales with which the post-burst/glitch timing and spectral
properties will recover to those in quiescence hold a great potentiality of
better understanding the mechanisms which rule the outbursts of AXPs.  In
particular, the comparison with the \xte\ outburst evolution might help us in
understanding whether other parameters play an important role in the observed
properties.

Finally, we note the the BAT detection of the bursts from \cxou\ opens new
prospectives for detecting further burst from known AXPs and for identifying new
AXPs/SGRs with the \swift\ mission.

\begin{theacknowledgments}
The complete list of the co-authors is: S. Campana,
A. Corongiu, J. Cummings, M. Dahlem, M. Falanga, P. Esposito, D. G\"otz, S.
Mereghetti, P. M. Muno, R. Perna, R. Turolla and S. Zane.
This work is partially supported at OAR through ASI, MIUR, and INAF grants. We
acknowledge financial contribution from contract ASI-INAF I/023/05/0.
\end{theacknowledgments}




\bibliography{sample}


\end{document}